\documentclass[aps,pra,reprint, superscriptaddress,showpacs]{revtex4-1}

\usepackage{amsmath,amssymb}
\usepackage{txfonts}
\usepackage[dvipdfmx]{graphicx}
\usepackage{subfigure}
\usepackage{url}
\begin{document}


\title{Quantum correlation in degenerate optical parametric oscillators with mutual injections}



\author{Kenta Takata}
\email[]{takata.kenta@lab.ntt.co.jp}
\affiliation{ImPACT, Japan Science and Technology Agency, Gobancho 7, Chiyoda-ku, Tokyo 102-0076, Japan}
\affiliation{National Institute of Informatics, Hitotsubashi 2-1-2, Chiyoda-ku, Tokyo 101-8403, Japan}
\affiliation{Department of Information and Communication Engineering, The University of Tokyo, Hongo 7-3-1, Bunkyo-ku, Tokyo 113-8654, Japan}

\author{Alireza Marandi}
\affiliation{ImPACT, Japan Science and Technology Agency, Gobancho 7, Chiyoda-ku, Tokyo 102-0076, Japan}
\affiliation{National Institute of Informatics, Hitotsubashi 2-1-2, Chiyoda-ku, Tokyo 101-8403, Japan}
\affiliation{E. L. Ginzton Laboratory, Stanford University, Stanford, California 94305, USA}

\author{Yoshihisa Yamamoto}
\affiliation{ImPACT, Japan Science and Technology Agency, Gobancho 7, Chiyoda-ku, Tokyo 102-0076, Japan}
\affiliation{National Institute of Informatics, Hitotsubashi 2-1-2, Chiyoda-ku, Tokyo 101-8403, Japan}
\affiliation{E. L. Ginzton Laboratory, Stanford University, Stanford, California 94305, USA}



\date{\today}

\begin{abstract}
We theoretically and numerically study the quantum dynamics of two degenerate optical parametric oscillators with mutual injections. The cavity mode in the optical coupling path between the two oscillator facets is explicitly considered. Stochastic equations for the oscillators and mutual injection path based on the positive $P$ representation are derived. The system of two gradually pumped oscillators with out-of-phase mutual injections is simulated, and its quantum state is investigated. When the incoherent loss of the oscillators other than the mutual injections is small, the squeezed quadratic amplitudes $\hat{p}$ in the oscillators are positively correlated near the oscillation threshold. It indicates finite quantum correlation, estimated via Gaussian quantum discord, and the entanglement between the intracavity subharmonic fields. When the loss in the injection path is low, each oscillator around the phase transition point forms macroscopic superposition even under a small pump noise. It suggests that the squeezed field stored in the low-loss injection path weakens the decoherence in the oscillators.
\end{abstract}




\maketitle

\section{INTRODUCTION}
Optics has contributed to the fundamental test and development of quantum mechanics from its dawn.
It is mainly because photons can have a large energy scale so that they are robust to thermal noise
and interact with matters via coherent multi-photon processes.
Especially, an optical parametric oscillator (OPO) \cite{paper:Byer75,paper:TBUL92} is a good tool to investigate quantum-mechanical effects,
since the photon pair generated in the phase-matched quadratic process \cite{paper:ABDP62} is correlated due to the energy and momentum conservation.
There has been vast theoretical and experimental literature on squeezing \cite{paper:Walls83,paper:CG84,paper:CW85,paper:WKHW86,paper:DDCR04,paper:TYYF07,paper:VMCH08}, entanglement \cite{paper:EPR35,paper:RD88,paper:Reid89,paper:RD89,paper:DR90,paper:OPKP92,paper:OD05,paper:VCCM05,paper:OO06,paper:JFBP06,paper:STJP06,paper:JWMT09} and quantum teleportation \cite{paper:BPME97,paper:BK98_2,paper:FSBA98} based on OPOs.
And now, applications of this device to state-of-the-art technologies such as frequency combs \cite{paper:WVB10,paper:LMBV11}, coherent feedback control \cite{paper:CTSA13}
and quantum information processing \cite{paper:YUAS13} is a new and important direction.

Inspired by the quantum simulators \cite{paper:FSJP08,paper:KCKI10,paper:SBMT11} and the concept of the bosonic system with artificial feedback \cite{paper:BYY11} to simulate the Ising spin system,
we have proposed and examined the systems called ``coherent Ising machines'' \cite{paper:UTY11,paper:TUY12,paper:WMWB13}, which are networks based on optical oscillators with mutual injections.
In these machines, each Ising spin is emulated by the field of each oscillator.
Also, the Ising Hamiltonian is mapped to the effective photonic loss of the whole system which is equivalent to the sum of the gain coefficients of all the oscillators.
When the machine is gradually pumped from below the oscillation threshold, it is expected that the system oscillates with the minimum gain balancing the loss, and the resulting state corresponds to the ground state of the simulated Hamiltonian.

We first have proposed the machine based on injection-locked lasers \cite{paper:UTY11,paper:TUY12}.
However, lasers usually do not have a fixed phase for their coherent fields to a certain reference,
thus the polarization or phase states of them cannot be binary as the Ising spins.
However, in the Ising machine based on degenerate OPOs (DOPOs) \cite{paper:WMWB13}, each DOPO above the threshold takes one of the binary phase states, thus they can be utilized to simulate the Ising spins.
Simulation results show that it can find the ground states of all the instances of the anti-ferromagnetic Ising model (the MAX-CUT problem) on cubic graphs with up to twenty spins. In an experimental demonstration \cite{paper:MWTB14}, a network of four OPOs successfully solved a problem of frustrated spins.

An open question is if such coherent Ising machines have quantum features, and exploit them for their computation. In the quantum computation framework, theoretical evidences promise computational powers beyond the rich of conventional digital computers \cite{proc:Shor94,paper:Grover97}. However, the previous theoretical models of the coherent Ising machines, introduced for benchmarking, are based on the semi-classical approach. Although such a model can deal with up to the quadratic squeezing process, a doubled-variable phase space representation is needed to consider further non-classical effects such as macroscopic superposition and entanglement \cite{paper:DDVF07}.

In this paper, we study a fully quantum mechanical model for the system of two DOPOs with mutual injections using the positive $P$ representation \cite{paper:DG80}. We extend the model for a single DOPO \cite{paper:DMW81} by adding the signal cavity mode in the mutual injection path between the facets of the two DOPOs. We recognize that the previous work on parametric oscillators \cite{paper:OD05,paper:OO06} and harmonic oscillators \cite{paper:PR08,paper:PR09} assumes local evanescent-type couplings. In our work, however, we consider a nonlocal and dissipative coupling via the full-quantum model, in which the auxiliary system operators, as well as the primary oscillator operators, are treated by the rigorous description. For the simulation, a series of stochastic differential equations for the signal modes are derived. Additionally, it is verified that the adiabatic elimination of the injection path results in the standard linear injection terms for the field variables. Numerical simulations on the gradually pumped system with out-of-phase mutual injections are performed. When the dissipation in the mutual injection path is moderate, the intracavity fields of the two DOPOs near the oscillation threshold can be entangled due to the quantum correlation between the squeezed quadrature amplitudes $\hat{p} = (\hat{a} - \hat{a}^{\dagger})/(2i)$, which is estimated via Gaussian quantum discord \cite{paper:GP10,paper:AD10}. When the whole system is sufficiently closed, the injection path stores a squeezed vacuum. In this case, the entanglement is destroyed by enhanced fluctuations in the anti-squeezed quadrature amplitude $\hat{x} = (\hat{a} + \hat{a}^{\dagger})/2$. However, the weak fringes of the distribution functions of $p$ in the transition indicate that macroscopic superposition of zero-phase and $\pi$-phase exists in the DOPOs even with a small nonlinear pump noise. This suggests that the squeezed vacuum stored in the mutual injection path protects the macroscopic superposition from decoherence \cite{paper:KW88,paper:MR95}.

This paper is organized as follows. In Sec. \ref{sec:Model2DOPOMI}, we describe the Ito stochastic differential equations (SDEs) for the field variables in the system and relate our model to the semi-classical model previously studied. In Sec. \ref{sec:SS2DOPOMI}, we give the simulation setting and review some ingredients for the simulated quantities and properties. In Secs. \ref{sec:SR2DOPOMI} and \ref{sec:DC2DOPOMI}, we discuss the simulation result and simulation schemes for this system. Sec. \ref{sec:CL2DOPOMI} concludes the paper.

\section{Theoretical Model}\label{sec:Model2DOPOMI}
\subsection{System overview}
\begin{figure}[htbp]
\begin{center}
\includegraphics[width=8cm]{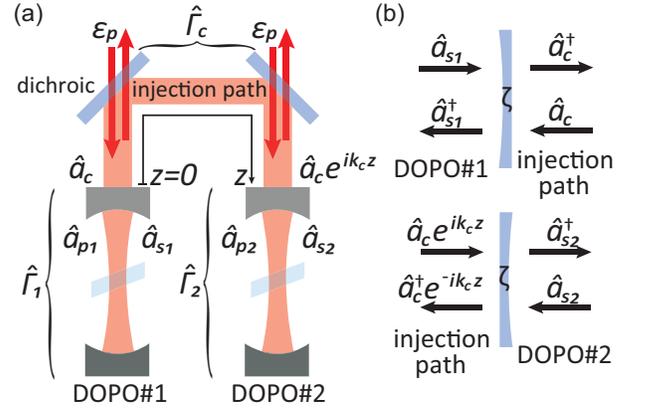}
\end{center}
\caption{(Color online) Schematic of the system. (a) The system comprises two DOPOs and a mutual injection path between them as a cavity. The two dichroic mirrors in the injection path are assumed to pass the pump field completely and highly reflect the signal field. (b) Beamsplitter interactions between the DOPO fields and the injection-path mode. The spatial phase of the injection-path field should be considered.}\label{fig:2DOPOSystem}
\end{figure}

Here, we describe the system considered in this study. Fig. \ref{fig:2DOPOSystem} (a) shows a schematic illustration of the system. It is composed of two DOPOs and a mutual injection path between them as a cavity. The two angled mirrors in the injection path are assumed to be dichroic. It means that they can highly reflect and confine the signal field of a frequency $\omega _s$ while entirely transmit the driving field with a frequency $\omega _d \sim \omega _p = 2 \omega _s$, where $\omega _p$ is the frequency of the pump mode. The coherent and real driving field $\varepsilon _p$ enters each DOPO to excite the pump mode. It is assumed to be classical and used as the phase reference. The bosonic annihilation and creation operators for the pump and signal modes in the DOPOs are denoted as $(\hat{a}_{pj}, \hat{a}_{pj}^{\dagger})$ and $(\hat{a}_{sj}, \hat{a}_{sj}^{\dagger})$, where $j = 1, 2$ is the index for the DOPOs.
Also, those for the signal mode in the injection path are written as $(\hat{a}_{c}, \hat{a}_{c}^{\dagger})$. Fig. \ref{fig:2DOPOSystem} (b) displays the coupling between the DOPO signal fields and the injection field described as beamsplitter interactions. Here, the field in the injection path interacts with the two DOPO fields at distant points, thus we have to consider the spatial phase explicitly. The phase factors for the bosonic operators at the facet of DOPO\#2 depend on the injection path length. When we set the $z$ axis as shown in \ref{fig:2DOPOSystem} (a), they are written as $\hat{a}_{c}\exp{(i k_c z)}$ and $\hat{a}_{c}^{\dagger}\exp{(-i k_c z)}$, where $k_c$ is the wave number for the signal mode. The mutual injections are in-phase couplings for $\exp{(i k_c z)} = \exp{(-i k_c z)} = 1$, and out-of-phase for $\exp{(i k_c z)} = \exp{(-i k_c z)} = -1$.

\subsection{System Hamiltonian}
The Hamiltonian for the system is an extension of that for a single DOPO \cite{paper:DMW81}, and can be written as 
\begin{equation}
{\cal H} = {\cal H}_{free} + {\cal H}_{int} + {\cal H}_{pump} + {\cal H}_{res} + {\cal H}_{BS},
\end{equation}
where the free Hamiltonian for the relevant modes is
\begin{equation}
{\cal H}_{free} = \sum _{j=1}^{2} \left( \hbar \omega _{p} \hat{a}_{pj}^{\dagger} \hat{a}_{pj} + \hbar \omega _{s} \hat{a}_{sj}^{\dagger} \hat{a}_{sj} \right) + \hbar \omega _{s} \hat{a}_{c}^{\dagger} \hat{a}_{c}.
\end{equation}
The quadratic nonlinear interaction Hamiltonian is
\begin{equation}
{\cal H}_{int} = i \hbar \sum _{j=1}^{2} \left[ \frac{\kappa}{2} \left( \hat{a}_{sj}^{\dagger \, 2} \hat{a}_{pj} - \hat{a}_{pj}^{\dagger} \hat{a}_{sj}^{2} \right) \right],
\end{equation}
where $\kappa$ denotes the coupling coefficient in terms of the subharmonic mode. The driving Hamiltonian is described by
\begin{equation}
{\cal H}_{pump} = i \hbar \sum _{j=1}^{2} \left[ \varepsilon _p \hat{a}_{pj}^{\dagger} \exp \left( - i \omega _d t \right) - \varepsilon _p \hat{a}_{pj} \exp \left( i \omega _d t \right)  \right],
\end{equation}
where the classical pumping flux $\varepsilon _p$ is set to be positive and real. The reservoir Hamiltonian for the signal and pump modes is written as
\begin{align}
{\cal H}_{res} = & \hbar \sum _{j=1}^{2} \left( \hat{a}_{sj} \hat{\Gamma}_{Rsj}^{\dagger} + \hat{\Gamma}_{Rsj} \hat{a}_{sj}^{\dagger} + \hat{a}_{pj} \hat{\Gamma}_{Rpj}^{\dagger} + \hat{\Gamma}_{Rpj} \hat{a}_{pj}^{\dagger} \right) \nonumber \\
&+ \left( \hat{a}_{c} \hat{\Gamma}_{Rc}^{\dagger} + \hat{\Gamma}_{Rc} \hat{a}_{c}^{\dagger} \right),
\end{align}
where $\hat{\Gamma}_{Rsj}$, $\hat{\Gamma}_{Rpj}$ and $\hat{\Gamma}_{Rc}$ are the heat bath operators for the signal, pump and signal mode in the injection path. Finally, the beamsplitter interaction Hamiltonian between the injection-path mode and DOPO signal modes is denoted by
\begin{equation}
{\cal H}_{BS} = i \hbar \zeta \left(\hat{a}_{c} \hat{a}_{s1}^{\dagger} - \hat{a}_{c}^{\dagger} \hat{a}_{s1} + \hat{a}_{s2} \hat{a}_{c}^{\dagger} e^{-i k_c z} - \hat{a}_{s2}^{\dagger} \hat{a}_{c} e^{i k_c z} \right),
\end{equation}
where $\zeta$ is the interaction coefficient. Note that ${\cal H}_{BS}$ considers a part of dissipation of the signal fields from the DOPO cavities as coherent transmissions and reflections.

\subsection{Stochastic equations}
With the standard technique to treat the thermal bath \cite{book:Carmichael02}, we have the master equation for the density operator of the system. Here, we neglect the thermal detuning term. Furthermore, we introduce the positive $P$ representation \cite{paper:DG80} for the five modes to expand the density operator with a well-behaved probability distribution function $P\left(\boldsymbol{\alpha}, \boldsymbol{\beta} \right)$ 
\begin{equation}
\hat{\rho} = \int P\left(\boldsymbol{\alpha}, \boldsymbol{\beta} \right) \frac{|{\boldsymbol \alpha} \rangle \langle {\boldsymbol \beta ^*}|}{ \langle \boldsymbol{\beta} ^* | \boldsymbol{\alpha} \rangle } d^{10} \boldsymbol{\alpha} d^{10} \, \boldsymbol{\beta} . \label{eq:PositiveP2DOPO}
\end{equation}
Here, $\boldsymbol{\alpha} =$ ($\alpha _{s1}$, $\alpha _{s2}$, $\alpha _{p1}$, $\alpha _{p2}$, $\alpha _{c}$)$^{\mathrm{T}}$ and $\boldsymbol{\beta} =$ ($\beta _{s1}$, $\beta _{s2}$, $\beta _{p1}$, $\beta _{p2}$, $\beta _{c}$)$^{\mathrm{T}}$ contain ten $c$-number variables to describe the state. $|{\boldsymbol \alpha} \rangle =$ $|\alpha _{s1}\rangle$ $|\alpha _{s2}\rangle$ $|\alpha _{p1}\rangle$ $|\alpha _{p2}\rangle$ $|\alpha _{c}\rangle$ and $\langle {\boldsymbol \beta ^*} | =$ $\langle \beta _{c}^* |$ $\langle \beta _{p2}^* |$ $\langle \beta _{p1}^* |$ $\langle \beta _{s2}^* |$ $\langle \beta _{s1}^* |$ are the coherent product states for the total system. The positive $P$ representation gives a positive and appropriately normalized distribution function for every quantum state. $\alpha _{X}$ and $\beta _{X}$ undergo statistically independent processes in probabilistic simulations while they are complex conjugate in average, i.e. $\langle \alpha _{X} \rangle = \langle \beta _{X} \rangle ^*$. Here, $X$ is the index for the cavity modes.

We substitute Eq. (\ref{eq:PositiveP2DOPO}) into the master equation and use the operator algebra \cite{paper:DG80} for the probability distribution description. After switching to the rotating frame with the driving frequency $\omega _d$ for the pump and $\omega _d /2$ for the signal mode, we obtain the Fokker-Planck equation (FPE) for the distribution $P\left(\boldsymbol{\alpha}, \boldsymbol{\beta} \right)$: \pagebreak
\begin{widetext}
\begin{align}
\frac{\partial}{\partial t} P\left(\boldsymbol{\alpha}, \boldsymbol{\beta} \right) 
= \Bigg\{ & \sum_{j=1}^{2} \Bigg[ \frac{\partial}{\partial \alpha _{sj}} \bigg( \left( \gamma _s + i \Delta _s \right) \alpha _{sj} - \kappa \beta _{sj} \alpha _{pj} \bigg) 
+ \frac{\partial}{\partial \beta _{sj}} \bigg( \left( \gamma _s - i \Delta _s \right) \beta _{sj} - \kappa \alpha _{sj} \beta _{pj} \bigg) 
+ \frac{\partial}{\partial \alpha _{pj}} \left( \left( \gamma _p + i \Delta _p \right) \alpha _{pj} - \varepsilon _p + \frac{\kappa^2}{2} {\alpha}_{sj}^{2} \right) \nonumber \\
& \qquad + \frac{\partial}{\partial \beta _{pj}} \left( \left( \gamma _p - i \Delta _p \right) \beta _{pj} - \varepsilon _p + \frac{\kappa^2}{2} {\beta}_{sj}^2 \right) 
+ \frac{1}{2} \left( \frac{\partial ^2}{\partial {\alpha} _{pj} ^2} \kappa \alpha _{pj}  + \frac{\partial ^2}{\partial {\beta} _{pj} ^2}  \kappa \beta _{pj}  + \frac{\partial ^2}{\partial \alpha _{sj} \partial \beta _{sj}} \Gamma _{sj}  + \frac{\partial ^2}{\partial \alpha _{pj} \partial \beta _{pj}} \Gamma _{pj}  \right) \Bigg] \nonumber \\
& \qquad + \Bigg[ \frac{\partial}{\partial \alpha _{c}} \left( \gamma _c + i \Delta _s \right) \alpha _{c} + \frac{\partial}{\partial \beta _{c}} \left( \gamma _c + i \Delta _s \right) \beta _{c} + \frac{1}{2}  \frac{\partial ^2}{\partial \alpha _{c} \partial \beta _{c}} \Gamma _{c} 
- \frac{\partial}{\partial \alpha _{s1}} \zeta \alpha _c -\frac{\partial}{\partial \beta _{s1}} \zeta \beta _c + \frac{\partial}{\partial \alpha _{s2}} \zeta \alpha _c e^{i \theta} + \frac{\partial}{\partial \beta _{s2}} \zeta \beta _c e^{-i \theta} \nonumber \\
& \qquad + \frac{\partial}{\partial \alpha _{c}} \zeta \left(\alpha _{s1} - \alpha _{s2} e^{-i \theta} \right) + \frac{\partial}{\partial \beta _{c}}  \zeta \left(\beta _{s1} - \beta _{s2} e^{i \theta} \right)
\Bigg]
\Bigg\}  P\left(\boldsymbol{\alpha}, \boldsymbol{\beta} \right), \label{eq:FPERF2DOPOs}
\end{align}
\end{widetext}
where $\theta = k_c z$, and $\Delta _s = \omega _s - \omega _d/2$ and $\Delta _p = \omega _p - \omega _d$ are the detuning between the cavity modes and the driving field. The components of the last two columns in Eq. (\ref{eq:FPERF2DOPOs}) come from the beamsplitter coupling. We may choose a more convenient reference for $\theta$ when the model is expanded for many-cavity systems.

With the Ito's rule \cite{book:Gardiner09} which gives the correspondence between the FPEs and SDEs, we reach a series of Ito SDEs for the $c$-number variables $\boldsymbol{\alpha}$ and $\boldsymbol{\beta}$
\begin{align}
d \left[ \begin{array}{cc}
\alpha _{s1} \\
\beta _{s1}
\end{array} \right] = &
\left[ \begin{array}{cc}
-\left(\gamma _s + i \Delta _s \right) \alpha _{s1} + \kappa \beta _{s1} \alpha _{p1}+ \zeta \alpha _c \\
-\left(\gamma _s - i \Delta _s \right) \beta _{s1} + \kappa \alpha _{s1} \beta _{p1} + \zeta \beta _c
\end{array} \right] dt \nonumber \\
& \qquad \qquad + \left[ \begin{array}{cc}
\kappa \alpha _{p1} & \Gamma _s \\
\Gamma _s & \kappa \beta _{p1}
\end{array} \right]^{1/2}
\left[ \begin{array}{cc}
d W_{\alpha s1}(t)\\
d W_{\beta s1}(t)
\end{array} \right], \label{eq:SDESignal1}
\end{align} 
\begin{align}
d \left[ \begin{array}{cc}
\alpha _{s2} \\
\beta _{s2}
\end{array} \right] = &
\left[ \begin{array}{cc}
-\left(\gamma _s + i \Delta _s \right) \alpha _{s2} + \kappa \beta _{s2} \alpha _{p2}- \zeta \alpha _c e^{i \theta} \\
-\left(\gamma _s - i \Delta _s \right) \beta _{s2} + \kappa \alpha _{s2} \beta _{p2} - \zeta \beta _c e^{-i \theta}
\end{array} \right] dt \nonumber \\
& \qquad \qquad + \left[ \begin{array}{cc}
\kappa \alpha _{p2} & \Gamma _s \\
\Gamma _s & \kappa \beta _{p2}
\end{array} \right]^{1/2}
\left[ \begin{array}{cc}
d W_{\alpha s2}(t)\\
d W_{\beta s2}(t)
\end{array} \right], \label{eq:SDESignal2}
\end{align} 
\begin{align}
d \left[ \begin{array}{cc}
\alpha _{p1} \\
\beta _{p1}
\end{array} \right] = &
\left[ \begin{array}{cc}
\displaystyle \varepsilon _p -\left(\gamma _p + i \Delta _p \right) \alpha _{p1} - \frac{\kappa}{2} \alpha _{s1} ^2 \\
\,  \\
\displaystyle \varepsilon _p -\left(\gamma _p - i \Delta _p \right) \beta _{p1} - \frac{\kappa}{2} \beta _{s1} ^2
\end{array} \right] dt \nonumber \\
& \qquad \qquad + \left[ \begin{array}{cc}
0 & \Gamma _p \\
\Gamma _p & 0
\end{array} \right]^{1/2}
\left[ \begin{array}{cc}
d W_{\alpha p1}(t)\\
d W_{\beta p1}(t)
\end{array} \right], \label{eq:SDEPump1}
\end{align} 
\begin{align}
d \left[ \begin{array}{cc}
\alpha _{p2} \\
\beta _{p2}
\end{array} \right] = &
\left[ \begin{array}{cc}
\displaystyle \varepsilon _p -\left(\gamma _p + i \Delta _p \right) \alpha _{p2} - \frac{\kappa}{2} \alpha _{s2} ^2 \\
\,  \\
\displaystyle \varepsilon _p -\left(\gamma _p - i \Delta _p \right) \beta _{p2} - \frac{\kappa}{2} \beta _{s2} ^2
\end{array} \right] dt \nonumber \\
& \qquad \qquad + \left[ \begin{array}{cc}
0 & \Gamma _p \\
\Gamma _p & 0
\end{array} \right]^{1/2}
\left[ \begin{array}{cc}
d W_{\alpha p2}(t)\\
d W_{\beta p2}(t)
\end{array} \right], \label{eq:SDEPump2}
\end{align} 
\begin{align}
d \left[ \begin{array}{cc}
\alpha _{c} \\
\beta _{c}
\end{array} \right] = &
\left[ \begin{array}{cc}
\displaystyle -\left(\gamma _c + i \Delta _s \right) \alpha _{c} - \zeta \alpha _{s1} + \zeta \alpha _{s2} e^{i \theta} \\
\,  \\
\displaystyle -\left(\gamma _c - i \Delta _s \right) \beta _{c} - \zeta \beta _{s1} + \zeta \beta _{s2} e^{-i \theta}
\end{array} \right] dt \nonumber \\
& \qquad \qquad + \left[ \begin{array}{cc}
0 & \Gamma _c \\
\Gamma _c & 0
\end{array} \right]^{1/2} 
\left[ \begin{array}{cc}
d W_{\alpha c}(t)\\
d W_{\beta c}(t)
\end{array} \right], \label{eq:SDECenter}
\end{align}
where $d W_{X}(t)$ is the real Wiener increment statistically independent of each other. This corresponds to the noise term in the equivalent Langevin equation whose autocorrelation is a delta function. Adding oscillators and injection paths is straightforward, thus this model will give a fully quantum mechanical treatment of mutually injecting oscillator networks.

Here, we consider the case of a resonant driving $\Delta _s = \Delta _p = 0$ and zero temperature $\Gamma _s = \Gamma _p = \Gamma _c = 0$. In addition, we adiabatically eliminate the pump variables with an assumption that the pump fields decay sufficiently faster than the signal fields. We can take diagonal diffusion amplitude matrices for the DOPO signal fields, then we have a simplified model as follows
\begin{align}
d \alpha _{s1} = & \bigg[ -\gamma _s \alpha _{s1} + \frac{\kappa}{\gamma _p}\Big( \varepsilon _p - \frac{\kappa}{2} \alpha _{s1}^2 \Big) \beta _{s1} + \zeta \alpha _c \bigg] dt \nonumber \\
& \qquad \qquad \qquad + \sqrt{\frac{\kappa}{\gamma _p} \left( \varepsilon _p - \frac{\kappa}{2} \alpha _{s1}^2 \right)} \, d W_{\alpha s1}(t), \label{eq:SDEas1AE}
\end{align}
\begin{align}
d \beta _{s1} = & \bigg[ -\gamma _s  \beta _{s1} + \frac{\kappa}{\gamma _p} \Big( \varepsilon _p - \frac{\kappa}{2} \beta _{s1}^2 \Big) \alpha _{s1}+ \zeta \beta _c \bigg] dt  \nonumber \\
& \qquad \qquad \qquad  + \sqrt{\frac{\kappa}{\gamma _p} \left( \varepsilon _p - \frac{\kappa}{2} \beta _{s1}^2 \right)} \, d W_{\beta s1}(t),
\end{align}
\begin{align}
d \alpha _{s2} = & \bigg[ -\gamma _s \alpha _{s2} + \frac{\kappa}{\gamma _p}\Big( \varepsilon _p - \frac{\kappa}{2} \alpha _{s2}^2 \Big) \beta _{s2} - \zeta \alpha _c e^{i \theta} \bigg] dt \nonumber \\
& \qquad \qquad \qquad  + \sqrt{\frac{\kappa}{\gamma _p}\left( \varepsilon _p - \frac{\kappa}{2} \alpha _{s2}^2 \right)} \, dW_{\alpha s2}(t),
\end{align}
\begin{align}
d \beta _{s2} = & \bigg[ - \gamma _s  \beta _{s2} + \frac{\kappa}{\gamma _p} \Big( \varepsilon _p - \frac{\kappa}{2} \beta _{s2}^2 \Big) \alpha _{s2} - \zeta \beta _c e^{-i \theta} \bigg] dt \nonumber \\
& \qquad \qquad \qquad  + \sqrt{\frac{\kappa}{\gamma _p} \left( \varepsilon _p - \frac{\kappa}{2} \beta _{s2}^2 \right)} \, dW_{\beta s2}(t), \label{eq:SDEbs2AE}
\end{align}
\begin{align}
d \alpha _{c} &= \left( -\gamma _c \alpha _{c} - \zeta \alpha _{s1} + \zeta \alpha _{s2} e^{i \theta} \right) dt, \\
d \beta _{c} &= \left( -\gamma _c  \beta _{c} - \zeta \beta _{s1} + \zeta \beta _{s2} e^{-i \theta}  \right) dt. \label{eq:SDEbcAE}
\end{align}

\subsection{Adiabatic elimination of the injection path}
To see the correspondence between our model and one for cavity systems with coherent external injections (e.g. Ref. \citenum{paper:WMWB13}), we further consider the limit where the injection-path mode is adiabatically eliminated, i.e. $\gamma _c \gg \gamma _s$. The field in the injection path at the steady state is given by
\begin{eqnarray}
\alpha _c ^ {ss} &=& \frac{1}{\gamma _c} \left( - \zeta \alpha _{s1} + e^{i \theta} \zeta \alpha _{s2} \right), \label{eq:acss} \\
\beta _c ^ {ss} &=& \frac{1}{\gamma _c} \left( - \zeta \beta _{s1} + e^{-i \theta} \zeta \beta _{s2} \right). \label{eq:bcss}
\end{eqnarray}
At the stable steady state under the phase locking due to the mutual injections, $\left(\langle \alpha _{s1} \rangle, \langle \beta _{s1} \rangle \right) = \left(\langle \alpha _{s2} \rangle e^{i \theta},\langle \beta _{s2} \rangle e^{-i \theta} \right)$ is expected. Hence, the injection path will be empty there, i.e. $\langle \alpha _c ^ {ss} \rangle = \langle \beta _c ^ {ss} \rangle = 0$.

Substituting Eqs. (\ref{eq:acss}) and (\ref{eq:bcss}) into (\ref{eq:SDEas1AE}) - (\ref{eq:SDEbs2AE}), we have the SDEs for the intracavity signal fields with adiabatic elimination of the pump and the injection path. When we further define the effective signal loss $\gamma' _s$ and the normalized beamsplitter coupling $\xi$ as
\begin{equation}
\gamma' _s = \gamma _s + \frac{\zeta ^2}{\gamma _c}, \quad \xi = \frac{\zeta ^2}{\gamma _s \gamma _c + \zeta ^2}, \label{eq:gammaps_xi}
\end{equation}
then we have the normalized SDEs for the signal modes with the eliminated injection path as
\begin{align}
&d \eta _j = \left[ - \eta _j + \mu _j \left( \lambda - \eta _j ^2 \right) + \xi \eta _k e^{i \theta} \right] d \tau + g \sqrt{\lambda - \eta _j ^2} d W_{\eta j}(\tau), \label{eq:2DOPOSDENM1} \\
&d \mu _j = \left[ - \mu _j + \eta _j \left( \lambda - \mu _j ^2 \right) + \xi \mu _k e^{-i \theta} \right] d \tau + g \sqrt{\lambda - \mu _j ^2} d W_{\mu _j}(\tau). \label{eq:2DOPOSDENM2}
\end{align}
Here, $\eta _j = g \alpha _{sj}$, $\mu _j = g \beta _{sj}$ and $g = \kappa / \sqrt{2 \gamma'_s \gamma _p}$ is the normalized parametric gain coefficient serving as a noise parameter. $\lambda = \varepsilon _p / \varepsilon _{th}$ is the normalized pumping rate and $\varepsilon _{th} = \gamma' _s \gamma _p/\kappa$ is the classical oscillation threshold. The time is scaled with the signal cavity lifetime, i.e. $\tau = \gamma' _s t$. $d W_{\eta j}(\tau)$ and $d W_{\mu _j}(\tau)$ are rescaled Wiener increments. The linear mutual injection terms with a explicit coupling phase $\xi \eta _k e^{i \theta}$ and $\xi \mu _k e^{-i \theta}$ have the same forms as those introduced in the semi-classical model \cite{paper:WMWB13}. Therefore, the theoretical framework studied here guarantees the validity of the standard injection model also in the quantum mechanical phase space representations, if the dynamics in the injection path can be neglected.

\section{Simulation Setting}\label{sec:SS2DOPOMI}
\subsection{Working equations}
Here, we discuss and review other elements important for the simulation. First, we describe the simulation setting. In this study, we focus on the out-of-phase mutual injections, namely $e^{i \theta} = e^{-i \theta} = -1$, expecting the out-of-phase correlation between the two macroscopic DOPO fields. We normalize Eqs. (\ref{eq:SDEas1AE}) - (\ref{eq:SDEbcAE}) as
\begin{align}
d \eta _j & = \left[ - \gamma _{sn} \eta _j + \mu _j \left( \lambda - \eta _j ^2 \right) + \zeta _n \eta _c \right] d \tau + g \sqrt{\lambda - \eta _j ^2} d W_{\eta j}(\tau), \label{eq:2DOPOSDEFullNM1} \\
d \mu _j & = \left[ - \gamma _{sn} \mu _j + \eta _j \left( \lambda - \mu _j ^2 \right) + \zeta _n \mu _c \right] d \tau + g \sqrt{\lambda - \mu _j ^2} d W_{\mu _j}(\tau). \label{eq:2DOPOSDEFullNM2} \\
& \qquad d \eta _c = \left( - \gamma _{cn} \eta _c - \zeta _n \eta _1 - \zeta _n \eta _2 \right) d \tau, \label{eq:2DOPOSDEFullNM3} \\
& \qquad d \mu _c = \left( - \gamma _{cn} \mu _c - \zeta _n \mu _1 - \zeta _n \mu _2 \right) d \tau, \label{eq:2DOPOSDEFullNM4}
\end{align}
where
\begin{align}
\gamma _{sn} &= \frac{\gamma _s}{\gamma' _s} = \frac{\gamma _s \gamma _c}{\gamma _s \gamma _c +\zeta ^2},\\
\gamma _{cn} &= \frac{\gamma _c}{\gamma' _s} = \frac{\gamma _c ^2}{\gamma _s \gamma _c +\zeta ^2},\\
\zeta _{n} &= \frac{\zeta}{\gamma' _s} = \frac{\zeta \gamma _c}{\gamma _s \gamma _c +\zeta ^2},
\end{align}
and simulate Eqs. (\ref{eq:2DOPOSDEFullNM1}) - (\ref{eq:2DOPOSDEFullNM4}) hereafter. The time unit in all the results is the effective cavity lifetime $1/\gamma' _s$.

Here, we refer again to the fact that the beamsplitter coupling in this model can explicitly take into account a large part of \textit{dissipation} from the DOPO cavities. The rest incoherent decay, which may phenomenologically include absorption and scatting in the oscillator, is considered by the conventional parameter $\gamma _s$. For high-$Q$ oscillators, we can expect the case $\zeta > \gamma _s$, and this is an important condition for the system to show non-trivial quantum effects. We set $\zeta = 1$, and $\gamma _s$ and $\gamma _c$ can be smaller in the simulations.

The noise parameter $g = \kappa / \sqrt{2 \gamma'_s \gamma _p}$ determines the typical order of the photon number inside the DOPOs above the oscillation threshold. Basically, we focus on the case of potentially larger photon numbers from the practical point of view, thus fix this parameter as $g \sim 0.01$. This gives $1/g^2 \sim 10000$ photons in the DOPOs at oscillation. For different decay parameters $\gamma'_s$ and $\gamma _p$, the nonlinearity $\kappa$ is changed accordingly to keep the value of $g$.

\subsection{Observable moments and distribution functions}
Drummond and Gardiner \cite{paper:DG80} have shown that normally ordered moments of the single mode oscillator can be obtained by the expectation value of corresponding $c$-number products. The trivial extension to the two-mode case with the commutability of bosonic operators for spatially separate modes indicates
\begin{equation}
\langle \hat{a}^{\dagger \, j}_{s1} \hat{a}^{\dagger \, k}_{s2} \hat{a}_{s1}^l \hat{a}_{s1}^m \rangle  = \int \beta _{s1}^j \beta _{s2}^k \alpha _{s1} ^l \alpha _{s2} ^m \, P\left( \{\alpha \} , \{\beta \} \right) \, d^4 \boldsymbol{\alpha} _{s}d^4 \boldsymbol{\beta} _{s}. \label{eq:Moments2DOPO}
\end{equation}
Here, $\{\alpha \} = \{\alpha _{s1}, \alpha _{s2}\}$, $\{\beta \} = \{\beta _{s1}, \beta _{s2}\}$ and other irrelevant modes are traced out. We consider and simulate the moments up to second order including ones with different modes, by unrestricted sampling Monte Carlo integration. 

In this study, we define the quadrature amplitudes in the DOPOs as $\hat{x}_j = (\hat{a}_j + \hat{a}_j^{\dagger})/2$ and $\hat{p}_j = (\hat{a}_j - \hat{a}_j^{\dagger})/(2 i)$. The distribution functions \cite{paper:YS86} of them are given by the diagonal element of the density operator for the corresponding eigenstates
\begin{align}
& P(x_j) = \langle x_j | \hat{\rho}_{sj} | x_j \rangle = \int P(\alpha _{sj}, \beta _{sj}) \frac{\langle x_j | \alpha_{sj} \rangle \langle \beta _{sj}^* | x_j \rangle}{\langle \beta _{sj}^* | \alpha _{sj} \rangle} d^{2}\alpha _{sj} d^{2} \beta _{sj} \nonumber \\
&= \sqrt{\frac{2}{\pi}} \int P(\alpha _{sj}, \beta _{sj}) \, e^{- 2 x_j^2 + 2 x_j ( \alpha _{sj} + \beta _{sj} ) - ( \alpha _{sj} + \beta _{sj} )^2 /2 } \, d^{2}\alpha _{sj} d^{2} \beta _{sj},
\label{eq:DistxDOPOj}
\end{align}
\begin{align}
& P(p_j) = \langle p_j | \hat{\rho}_{sj} | p_j \rangle = \int P(\alpha _{sj}, \beta _{sj}) \frac{\langle p_j | \alpha_{sj} \rangle \langle \beta _{sj}^* | p_j \rangle}{\langle \beta _{sj}^* | \alpha _{sj} \rangle} d^{2}\alpha _{sj} d^{2} \beta _{sj} \nonumber \\
& = \sqrt{\frac{2}{\pi}} \int P(\alpha _{sj}, \beta _{sj}) \, e^{- 2 p_j^2 - i 2 p_j ( \alpha _{sj} - \beta _{sj} ) + ( \alpha _{sj} - \beta _{sj} )^2 /2 } \, d^{2}\alpha _{sj} d^{2} \beta _{sj}.
\label{eq:DistpDOPOj}
\end{align}
Here, $\hat{\rho}_{sj}$ is the partial density operator for the signal field in DOPO\#$j$, with the other states traced out. Also, we have used the inner products involving the eigenstates and coherent state
\begin{align}
\langle x_j | \alpha_{sj} \rangle &= \left(\frac{\pi}{2}\right) ^{-\frac{1}{4}} \exp \left(-2 x_j^2 + 2 x_j \alpha _{sj} - \frac{\alpha _{sj}^2}{2} - \frac{|\alpha _{sj}|^2}{2} \right), \\
\langle p_j | \alpha_{sj} \rangle &= \left(\frac{\pi}{2}\right) ^{-\frac{1}{4}} \exp \left(-2 p_j^2 - i 2 p_j \alpha _{sj} + \frac{\alpha _{sj}^2}{2} - \frac{|\alpha _{sj}|^2}{2} \right), \\
\langle \beta _{sj}^* | \alpha_{sj} \rangle &= \exp \left( - \frac{|\alpha _{sj}|^2}{2} - \frac{|\beta _{sj}|^2}{2} + \alpha _{sj} \beta _{sj} \right).
\end{align}
It is known that an oscillation in $P(p_j)$ is the evidence for existence of superposition components of coherent states \cite{paper:RY92}. In Eq. (\ref{eq:DistpDOPOj}), we see that in the manifold called \textit{classical subspace} where $\alpha _{sj}$ and $\beta _{sj}$ are real, the oscillation in $P(p_j)$ comes from the integration of the component $\exp \left[- i 2 p_j \left( \alpha _{sj} - \beta _{sj} \right) \right]$. Here, the quantum noise causing stochastic discrepancies between $\alpha _{sj}$ and $\beta _{sj}$ is found to be essential for observing the fringes. Also, in the other phase space representations where effectively $\beta = \alpha ^*$, $P(p_j)$ does not show any fringe due to the real exponent in the integrand.

\subsection{Criterion for entanglement}
To examine the entanglement between two intracavity signal fields, we adopt the criterion proposed by Duan \textit{et al} \cite{paper:DGCZ00}. Here, we consider the pair of Einstein-Podolsky-Rosen (EPR)-type operators \cite{paper:EPR35} $\hat{u} _{+} = \hat{x}_1 + \hat{x}_2$, $\hat{v} _{-} = \hat{p}_1 - \hat{p}_2$ and their fluctuation operators $\Delta \hat{u} _{+} = \hat{u} _{+} - \langle \hat{u} _{+} \rangle$ and $\Delta \hat{v} _{-} = \hat{v} _{-} - \langle \hat{v} _{-} \rangle$. The quadrature amplitudes defined here satisfy the commutation relation $\left[\hat{x}_j, \hat{p}_k \right] = i \delta _{jk}/2$. Thus, the condition for the entanglement (inseparability) between the two DOPO signal fields is given by
\begin{equation}
\langle \Delta \hat{u}_{+}^2 \rangle + \langle \Delta \hat{v}_{-}^2 \rangle < 1. \label{eq:ETGCriterion2DOPO}
\end{equation}

\subsection{Quantum discord}
Finally, we refer to quantum correlation estimated in this study. It is referred to as the property of a composite system which a local measurement changes the state of the whole system. It is a weaker but broader characteristic than entanglement, showing that even separable states can have some quantum features such as lack of complete distinguishability due to a nonorthogonal basis, and pure quantumness of each partial system.

Quantum discord \cite{paper:OZ01} is a measure of quantum correlation, based on two different ways to describe the mutual information of a bipartite system. Suppose we have a system $AB$ composed of partial systems $A$ and $B$. The mutual information based on the total system entropy is
\begin{equation}
I(\hat{\rho}_{AB}) = S(\hat{\rho}_{A}) + S(\hat{\rho}_{B}) -S(\hat{\rho}_{AB}), 
\end{equation}
where $S(\hat{\rho}) = - \mathrm{Tr}\left(\hat{\rho} \log \hat{\rho} \right)$ is the von Neumann entropy. On the other hand, the mutual information based on the conditional entropy $S(A|B)$ varies with the measurement basis for $B$, because the local measurement can perturb the total system. To evaluate the genuine quantum correlation, the measurement basis which disturbs the system least is chosen. The conditional mutual information in quantum theory is hence defined as
\begin{equation}
J^{\leftarrow }(\hat{\rho}_{AB}) = S(\hat{\rho}_{A}) - \inf_{\Pi _i^B} \sum_i p_i S(\hat{\rho}_{A|i}), \label{eq:MIConditional}
\end{equation}
where $i$ is the index for the components of the POVM measurement basis $\{\Pi _i^B \}$ for $B$. $\hat{\rho}_{A|i}$ is the posterior state of $A$ provided that the $i$th state is measured at $B$. The quantum discord is defined as the difference of them
\begin{equation}
D^{\leftarrow}(\hat{\rho}_{AB}) = I(\hat{\rho}_{AB}) - J^{\leftarrow }(\hat{\rho}_{AB}). \label{eq:QDiscordDef}
\end{equation}
A bipartite system with a finite discord surely has quantum correlation between its elements. A system without entanglement can have nonzero discord, and it has been reported that such a ``dirty'' state may be available for a nontrivial speedup in certain problems \cite{paper:LBAW08} with a quantum computing model called \textit{DQC1} \cite{paper:KL98}.

In general, the optimization about the measurement basis in Eq. (\ref{eq:MIConditional}) is hard. Its time complexity has been shown to be NP-complete for qubit systems \cite{paper:Huang14}, and been an open problem for continuous-variable states. However, for the case of Gaussian states and local measurement limited to Gaussian POVMs, analytic formulae \cite{paper:GP10,paper:AD10} for the discord have been derived. Also, it has been shown that they quantify the amount of genuine quantum correlation for a large part of Gaussian states, including two-mode squeezed states, coherent states and the vacuum state \cite{paper:PSBCL14}. In this study, we approximate the signal field in each oscillator as an Gaussian state and estimate the discord of them as this \textit{Gaussian quantum discord}. Here, we consider the unnormalized quadrature amplitudes for the two modes $\left[ \hat{r} \right] = 2 \left[\hat{x}_1, \hat{p}_1, \hat{x}_2, \hat{p}_2\right]$.
Then, a two-mode Gaussian state is characterized with the covariance matrix of them
\begin{equation}
\boldsymbol{\sigma} _G = \left[ \frac{1}{2} \langle \hat{r}_j \hat{r}_k + \hat{r}_k \hat{r}_j \rangle - \langle \hat{r}_j \rangle \langle \hat{r}_k \rangle  \right]
= \left( \begin{array}{cc}
\boldsymbol{\alpha}_M & \boldsymbol{\gamma}_M \\
\boldsymbol{\gamma}_M^{\mathrm{T}} & \boldsymbol{\beta}_M
\end{array} \right).
\end{equation}
where $\boldsymbol{\alpha}_M$, $\boldsymbol{\beta}_M$ and $\boldsymbol{\gamma}_M$ are $2 \times 2$ matrices. The state can also be equivalently featured by the quantities called \textit{symplectic invariants} defined as
\begin{equation}
A_{S} = \mathrm{det} \boldsymbol{\alpha}_M, \, B_{S} = \mathrm{det} \boldsymbol{\beta}_M, \, C_{S} = \mathrm{det} \boldsymbol{\gamma}_M, \, D_{S} = \mathrm{det} \boldsymbol{\sigma} _G.
\end{equation}
When we write the binary entropy function as $f_B (X) = \left( X + \frac{1}{2} \right) \log \left( X + \frac{1}{2} \right) - \left( X - \frac{1}{2} \right) \log \left( X - \frac{1}{2} \right)$, and the quantities called \textit{symplectic eigenvalues} as $\nu _{\pm}^2 = \frac{1}{2} \left( \Delta \pm \sqrt{\Delta^2 - 4 D_{S}} \right)$, $\Delta = A_{S} + B_{S} + 2 C_{S}$, the Gaussian quantum discord is given by
\begin{equation}
D^{\leftarrow}(\boldsymbol{\sigma} _G) = f_B \left(\sqrt{B_{S}} \right) - f_B (\nu _{-}) - f_B (\nu _{+}) + \inf _{\boldsymbol{\sigma} _0} f_B \left(\sqrt{\mathrm{det}\, \boldsymbol{\epsilon}} \right). \label{eq:GaussianDiscordDef}
\end{equation}
Here, $\boldsymbol{\sigma}_0$ is the measurement basis for the partial system $B$. $\boldsymbol{\epsilon}$ is the covariance matrix for the partial system $A$ after $B$ has been locally measured. The last term in Eq. (\ref{eq:GaussianDiscordDef}) can be optimized analytically within the range of Gaussian POVMs (adding Gaussian ancilla bits, symplectic transformations and a homodyne detection), yielding \cite{paper:AD10}
\begin{align}
&\inf _{\boldsymbol{\sigma} _0} \mathrm{det} \, \boldsymbol{\epsilon} = \nonumber \\
&\frac{2C_{S}^2 + \left(B_{S}-1\right)\left(D_{S}-A_{S}\right)+2|C_{S}|\sqrt{C_{S}^2+\left(B_{S}-1\right)\left(D_{S}-A_{S}\right) } }{\left(-1 + B_{S}\right)^2} \nonumber \\
& \textrm{if} \quad \left(D_{S} - A_{S}B_{S}\right)^2 \le \left(1+B_{S}\right)C_{S}^2 \left(A_{S}+D_{S}\right), \nonumber \\
\nonumber \\
&\frac{A_{S}B_{S}-C_{S}^2+D_{S}-\sqrt{C_{S}^4+\left(D_{S}-A_{S}\right)^2-2C_{S}^2\left(A_{S}B_{S}+D_{S}\right)}}{2B_{S}} \nonumber \\
& \textrm{otherwise}. \label{eq:GauusianDC1}
\end{align}
In addition, a simpler formula for two-mode squeezed thermal states (including squeezed vacua) has been also derived as \cite{paper:GP10}
\begin{equation}
\sqrt{\mathrm{det} \, \boldsymbol{\epsilon}} = \frac{\sqrt{A_{S}}+ 2 \sqrt{A_{S}B_{S}} + 2 C_{S}}{1 + \sqrt{B_{S}}}. \label{eq:GauusianDC2}
\end{equation}
A bipartite state with $D^{\leftarrow}(\boldsymbol{\sigma} _G) \ge 1$ always has entanglement between its elements. On the other hand, an entangled state can have a value of the quantum discord smaller than 1.

In our simulations, the smaller value in those calculated with Eqs. (\ref{eq:GaussianDiscordDef}), (\ref{eq:GauusianDC1}) and (\ref{eq:GauusianDC2}) is taken for each point to achieve a good approximation and avoid possible numerical instabilities, especially in the case of small pumping rates.

\section{Simulation Result}\label{sec:SR2DOPOMI}
Here, we show the results of the numerical simulation of the system with out-of-phase mutual injections. The initial state is fixed to the vacuum state, i.e. $\boldsymbol{\alpha} = \boldsymbol{\beta} = \boldsymbol{0}$. The system is gradually pumped \cite{paper:TY14}, meaning that the pumping rate $\lambda$ is slowly increased in time so that the DOPOs are continuously driven from the below to above of the oscillation threshold. This helps the DOPOs hold the state with the minimum effective gain and avoid being dragged into an unstable solution created by the mutual injection terms. we set the linear schedule for the pumping as
\begin{equation}
\lambda(t) = \frac{\lambda _f t}{t_f},
\end{equation}
where $\lambda _f$ and $t_f$ are the pump and time parameters for the final state. Note that the state is always transient because the pumping rate is continuously increased. Transient effects get clear when the decay rate in the injection path is small. However, the sweeping is sufficiently slow so that the DOPOs keep themselves stable. For the numerical integration on the SDEs, we adopt a second-order weak scheme \cite{book:BF07} originally proposed by Kloeden and Platen \cite{book:KP92}, with a time step $\Delta t = 2 \times 10^{-3}$.

We consider two cases here. In the first case, the signal intracavity decay rate $\gamma _s$ is varied under the condition that the loss in the injection path $\gamma _c$ is larger than it. Here, we hold $\gamma _c = 2 \gamma _s$, meaning that the loss of the whole system is increased with those. In the second case, we change only $\gamma _c$, keeping $\gamma _s$ small ($\gamma _s = 0.01$). Eqs. (\ref{eq:2DOPOSDEFullNM1}) - (\ref{eq:2DOPOSDEFullNM4}) are numerically integrated with fixed parameters $t_f = 200$, $\lambda _f = 1.5$, $\gamma _p = 100$, $\zeta = 1$ and $g \sim 0.01$ for these two. The results for the first and second cases are symbolized as (a) and (b) in the following figures, respectively. We take 50,000 stochastic trials to compute the observables. For the numerical stability, we introduce a boundary condition \cite{paper:KMR94} in the numerical algorithm to assure that the trajectories do not go out of the \textit{classical subspace} $|\eta _j| \le \sqrt{\lambda}$, $|\mu _j| \le \sqrt{\lambda}$. This is the case for a slowly pumped system where the injection path is nearly empty, because the steady solution of Eqs. (\ref{eq:2DOPOSDENM1}) and (\ref{eq:2DOPOSDENM2}) under the expected condition $\eta _j = \mu _j$, $\eta _2 = -\eta _1$ is in this manifold. The condition means that we neglect the second harmonic generation, the reverse process of parametric down-conversion, by the state out of the boundary.

\subsection{Mean photon number}
\begin{figure}[htbp]
\begin{center}
\includegraphics[width=8cm]{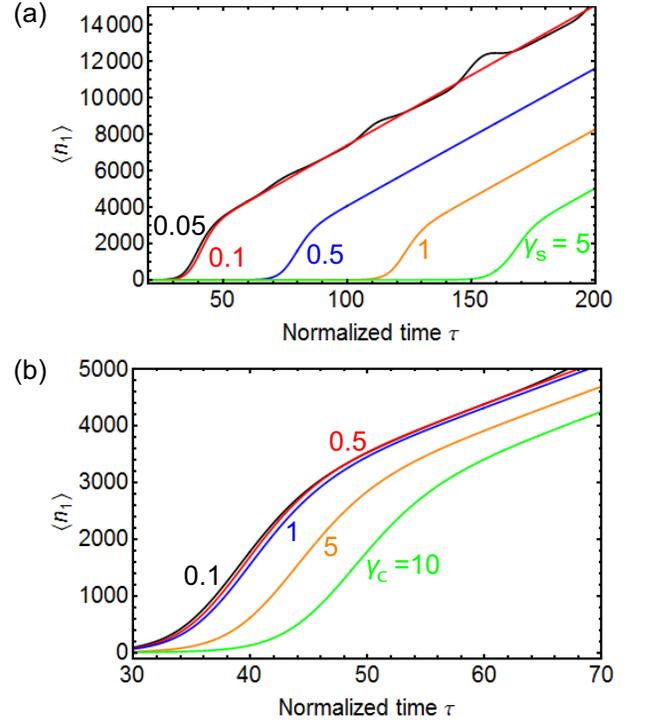}
\end{center}
\caption{(Color online) Transient of the intracavity photon number in a DOPO dependent on (a) the signal loss of the system under $\gamma _c = 2 \gamma _s$, (b) the decay rate in the injection path $\gamma _c$. The normalized oscillation threshold depends on the effective mutual injection strength, varying with the parameters. 50,000 stochastic runs for each curve.}\label{fig:PhotonN2DOPO}
\end{figure}
Fig. \ref{fig:PhotonN2DOPO} shows the transits of the mean photon number in the first DOPO. Those for the second oscillator are omitted because they look the same. The photon number rises, as the pumping rate increases linearly in time. The difference in the effective oscillation threshold and resulting intensity of the curves comes from the difference in the magnitude of the mutual injections. Eqs. (\ref{eq:gammaps_xi}), (\ref{eq:2DOPOSDENM1}) and (\ref{eq:2DOPOSDENM2}) are convenient to evaluate the threshold. When in the assumption that $\eta _1 = \mu _1 = - \eta _2 = - \mu _2$ and the field of the injection path is eliminated, the effective classical threshold is given by $\lambda _{th} = 1 - |\xi|$. For the cases of $\gamma _s =$ (0.05, 0.1, 0.5, 1, 5) in Fig. \ref{fig:PhotonN2DOPO} (a), the corresponding coupling coefficients and approximate thresholds are  $\xi =$ (0.995, 0.980, 0.67, 0.33, 0.020) and $\lambda _{th} =$ (0.0050, 0.020, 0.33, 0.67, 0.98), respectively. Also, for $\gamma _c =$ (0.1, 0.5, 1, 5, 10) in Fig. \ref{fig:PhotonN2DOPO} (b), $\xi =$ (0.999, 0.995, 0.990, 0.952, 0.909) and $\lambda _{th} =$ (0.0010, 0.0050, 0.0099, 0.048, 0.091). Note that $\lambda = (1.5/200) t$ for (a) and (b). A sufficiently closed system with mutual injections has a drastically smaller threshold than a single DOPO. At the same time, however, quantum noise in the system interrupts oscillation of the DOPOs and leads to less photons around the threshold than those classically expected by the extrapolation of the linear region. Also, the curve for $\gamma _s = 0.05$ in (a) shows a relaxation oscillation due to the injection path. Such a dynamics suggests a possibility that the variables escape the classical subspace. The line $\gamma _c = 0.1$ in (b) is also the case, although the oscillation is outside the range of the figure.

\subsection{Correlation function of quadrature amplitudes}
\begin{figure}[htbp]
\begin{center}
\includegraphics[width=8cm]{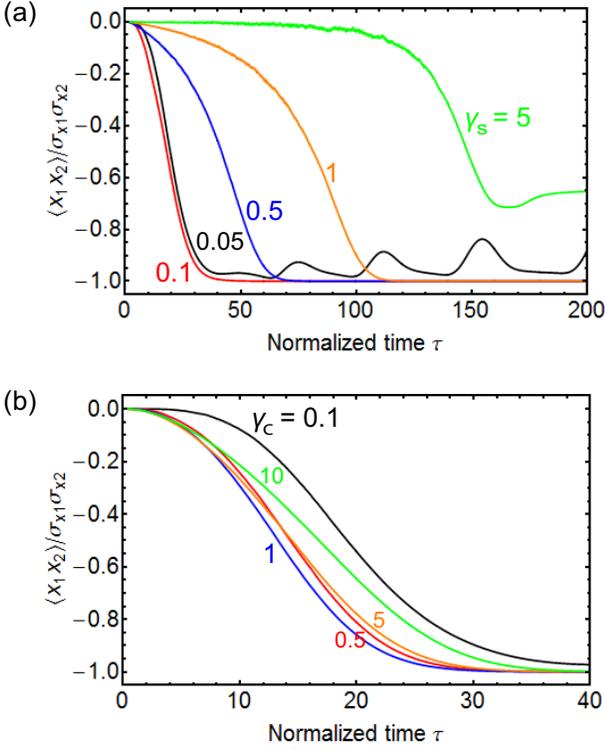}
\end{center}
\caption{(Color online) Time dependence of the correlation functions for the signal quadrature amplitude $x$. (a) The loss in the system with $\gamma _c = 2 \gamma _s$ and (b) only the loss in the injection path $\gamma _c$ are varied. $x_j$ are nagatively and macroscopically correlated along with the oscillation. 50,000 stochastic runs for each curve.} \label{fig:xCorrelation2DOPO}
\end{figure}
\begin{figure}[htbp]
\begin{center}
\includegraphics[width=8cm]{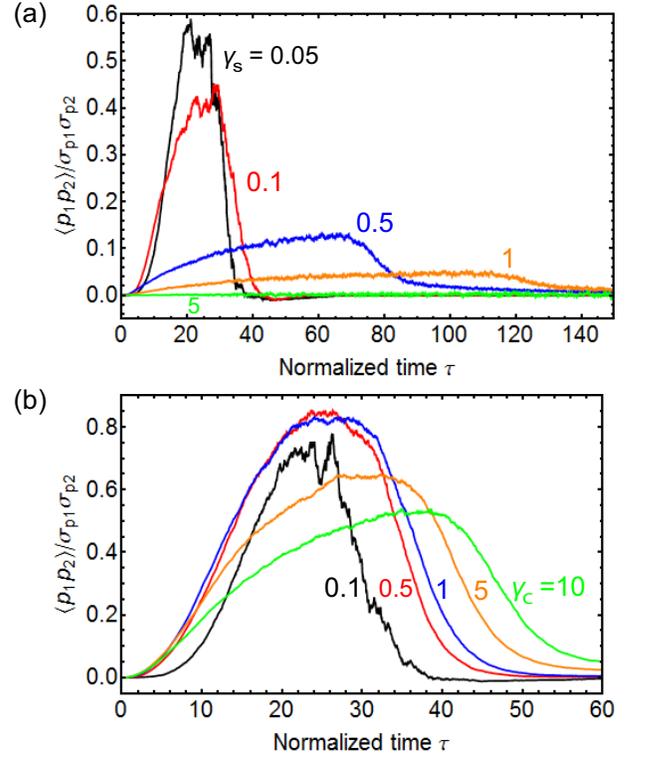}
\end{center}
\caption{(Color online) Time dependence of the correlation functions for the signal quadrature amplitude $p$. (a) The loss in the system with $\gamma _c = 2 \gamma _s$ and (b) only $\gamma _c$ are changed. $p_j$ are positively and microscopically correlated before the oscillation. 50,000 stochastic runs for each curve.} \label{fig:pCorrelation2DOPO}
\end{figure}
Fig. \ref{fig:xCorrelation2DOPO} and \ref{fig:pCorrelation2DOPO} displays the second order correlation functions for the quadrature amplitudes $x$ and $p$. They are normalized with the products of the standard deviations of the relevant amplitudes. In Fig. \ref{fig:xCorrelation2DOPO}, the negative correlation in $x_1$ and $x_2$ enhances as the pumping rate is increased and hence the photon number in the DOPOs rises. It means that the system gives the macroscopic out-of-phase order in $x$ due to the mutual injections, corresponding to the anti-ferromagnetic order of the most fundamental Ising model ${\cal \hat{H}} = \hat{\sigma}_{z1} \hat{\sigma}_{z2}$ programmed in the system. The relaxation oscillation is seen in the curve of $\langle \hat{x_1} \hat{x_2} \rangle$ for a small loss $\gamma _s$ of 0.05. Despite that the mutual injections negatively couple $\boldsymbol{\alpha}$ and $\boldsymbol{\beta}$, the curves of $\langle \hat{p_1} \hat{p_2} \rangle$ in Fig. \ref{fig:pCorrelation2DOPO} show that the instantaneous amplitudes $p_1$ and $p_2$ correlate positively. $\langle \hat{p_1} \hat{p_2} \rangle$ vanishes as the photon number rises and $\langle \hat{p_1} \rangle = \langle \hat{p_2} \rangle = 0$. Thus, it indicates that $\langle \hat{p_1} \hat{p_2} \rangle$ shows quantum correlation induced by the quantum noise and mutual injections. Opposite signs of correlation for the two quadrature amplitudes $x$ and $p$ suggests the entanglement of \textit{wavefunctions} \cite{paper:EPR35} of the two DOPO fields. We also got positive $\langle \hat{x_1} \hat{x_2} \rangle$ and negative $\langle \hat{p_1} \hat{p_2} \rangle$ in the case of in-phase mutual injections, corresponding to the artificial ferromagnetic interaction.

In Fig. \ref{fig:xCorrelation2DOPO} (b) and \ref{fig:pCorrelation2DOPO} (b), we see that a highly closed system ($\gamma _c = 0.1$) interrupts the formation of the correlation between the squeezed vacua in the DOPOs. Here, the injection path stores the squeezed vacuum from the DOPOs because $\zeta > \gamma _c$. Thus, it effectively works as an additional noise source to the DOPOs. $\gamma _c \approx \zeta$ gives good correlation in $p$, and the degradation of $\langle \hat{p_1} \hat{p_2} \rangle$ due to a larger $\gamma _c$ is not so significant. Note that broader peaks with $\gamma _c = 5$ and $10$ come from larger oscillation thresholds, as seen in Fig. \ref{fig:PhotonN2DOPO}.

\subsection{Total fluctuation of EPR-type operators}
\begin{figure}[htbp]
\begin{center}
\includegraphics[width=8cm]{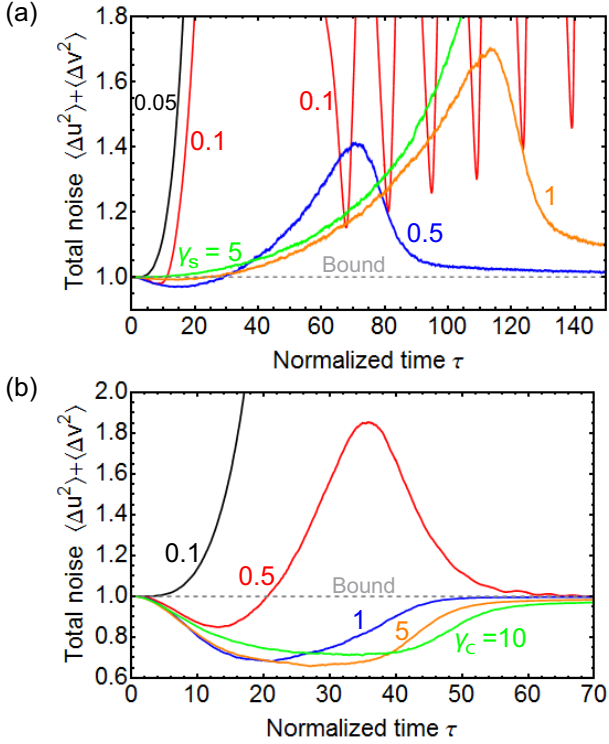}
\end{center}
\caption{(Color online) Time evolution of the total fluctuation in the EPR-type operators $\langle \Delta u_{+}^2 \rangle + \langle \Delta v_{-}^2 \rangle$ dependent on (a) the loss in the system with $\gamma _c = 2 \gamma _s$, (b) the loss in the injection path $\gamma _c$. $\langle \Delta u_{+}^2 \rangle + \langle \Delta v_{-}^2 \rangle < 1$ means the entanglement between the DOPO signal fields. 50000 stochastic runs for each curve.}\label{fig:TotalFluc2DOPO}
\end{figure}

Fig. \ref{fig:TotalFluc2DOPO} presents the total variance of the EPR-type operators varying with time. Here, $\langle \Delta u_{+}^2 \rangle + \langle \Delta v_{-}^2 \rangle < 1$ represents the entanglement between the DOPO signal fields. When the loss in the system is small, the total fluctuation sharply rises from the vacuum level in spite of a respectable correlation in $p$, seen in Fig. \ref{fig:pCorrelation2DOPO}. This means that the fluctuation of $\hat{v}_{-} = \hat{p}_1 - \hat{p}_2$ falls under the vacuum level of 0.5 before oscillation, while that of $\hat{u}_{+} = \hat{x}_1 + \hat{x}_2$ gets much larger than that due to the anti-squeezed noise field accumulated in the injection path. Both larger $\gamma _s$ and $\gamma _c$ denote more dissipation and less mutual injections, thus the curves in Fig. \ref{fig:TotalFluc2DOPO} (a) do not satisfy the entanglement criterion except for the small region up to $\tau \sim 30$. The oscillation in the curve for $\gamma _s = 0.1$ indicates that the phase locking of the coherent DOPO fields is hampered by the resonance of the injection path. This is also supported by the fluctuation in the quantum discord (Fig. \ref{fig:Discord2DOPO}(a)).

In Fig. \ref{fig:TotalFluc2DOPO} (b), when only $\gamma _c$ is increased the total noise comes to drop clearly below the bound before oscillation. Thus, the system has the entanglement there. This is because the damping of the field in the injection path gets faster while the system keeps a large amount of the mutual injections. Here, a good part of the output fields from the DOPOs coherently inject to each other. It is known that an entangled state cannot be produced only with local operations and classical communication (LOCC) \cite{paper:VPRK97}, thus the result here shows that the mutual injections can convey quantum information. As expected from the time range where the system shows the entanglement, it solely reflects the quantum correlation in $p$, i.e. $\langle \Delta v_{-}^2 \rangle < 0.5$. In fact, $\langle \Delta u_{+}^2 \rangle$ comes down only nearly to 0.5 (vacuum level) hence the total noise level is always larger than 0.5. This indicates that the entanglement is not perfect in the sense that the system shows classical correlation in $x$.

\subsection{Quantum discord}
\begin{figure}[htbp]
\begin{center}
\includegraphics[width=8cm]{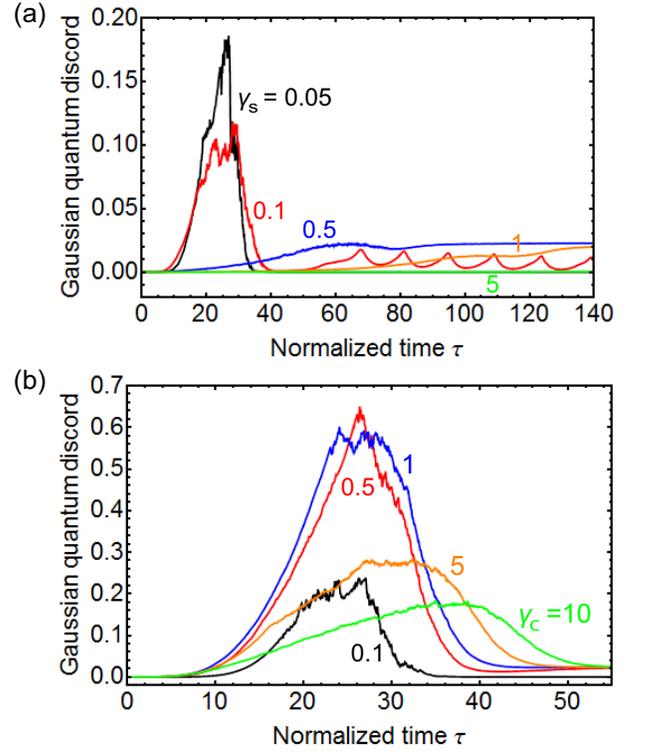}
\end{center}
\caption{(Color online) Quantum discord when the state is approximated as a bipartite Gaussian state. Squeezing in the DOPOs below the threshold and the mutual injections give a large discord. Coherent fields above the threshold in them and a coherent communication lead to a finite discord. 50000 stochastic runs for each curve.}\label{fig:Discord2DOPO}
\end{figure}
Fig. \ref{fig:Discord2DOPO} shows the approximate quantum discord when the state is considered as a Gaussian state. It basically reflects the quantum correlation in $p_j$ (see the correspondence between Fig. \ref{fig:pCorrelation2DOPO} and \ref{fig:Discord2DOPO}). When $p_j$ is squeezed and has positive correlation with that in the other DOPO, the system holds a relatively large discord. Moreover, many curves converge at a finite value $D^{\leftarrow} \sim 0.02$, except for the ones with $\gamma _s = 0.05, 1, 5$ in Fig. \ref{fig:Discord2DOPO} (a) and one with $\gamma _c = 0.1$ in (b). It is worth noting that this finite discord does not attribute to the squeezing in the DOPOs as previously discussed \cite{paper:GP10} for the case of squeezed thermal states, but the mixture of coherent states with classical correlation. We have found that the variances in $x_j$ and $p_j$ quickly verge on 1/4 after oscillation in the data here, thus the states there are well described as coherent states. Also, the curves with the finite discord give an almost perfect correlation of $x$ (injection-locking) at the final state such as $\langle \hat{x_1} \hat{x_2} \rangle =$ -0.9999 and -1.0000. On the other hand, a low-loss injection path and significant dissipation can cancel out the discord for the DOPOs well above the oscillation threshold. The lines without discord for $\gamma _s = 0.05, 5$ in (a) and one with $\gamma _c = 0.1$ in (b) have $\langle \hat{x_1} \hat{x_2} \rangle = $ -0.957, -0.6036 and -0.9615, respectively.

The ideal field state of the two DOPOs well above the threshold with the classical out-of-phase correlation and its covariance matrix are given by
\begin{equation}
\hat{\rho}_{cl} = \frac{1}{2} |\alpha _{cl}\rangle _1 \, |-\alpha _{cl}\rangle _2 \, _2 \langle -\alpha _{cl}| \, _1 \langle \alpha _{cl}| + \frac{1}{2} |-\alpha _{cl}\rangle _1 \, |\alpha _{cl}\rangle _2 \, _2 \langle \alpha _{cl}| \, _1 \langle -\alpha _{cl}|, \label{eq:DOforCLS}
\end{equation}
\begin{equation}
\boldsymbol{\sigma}(\hat{\rho}_{cl}) = \left(
\begin{array}{cccc}
4 \alpha _{cl}^2 + 1 & 0 & -4 \alpha _{cl}^2 & 0 \\
0 & 1 & 0 & 0 \\
-4 \alpha _{cl}^2 & 0 & 4 \alpha _{cl}^2 + 1 & 0 \\
0 & 0 & 0 & 1
\end{array}
\right), \label{eq:CVMforCLS}
\end{equation}
where $\alpha _{cl}$ is the real and positive amplitude of the coherent states in the DOPOs. We have found that the Gaussian discord calculated with Eq. (\ref{eq:CVMforCLS}) verges on $D^{\leftarrow} \sim 0.02356$ for $\alpha _{cl} \gtrsim 50$, which is in a good agreement with the values in the simulation. Eq. (\ref{eq:DOforCLS}) clearly represents a mixture of Gaussian states, thus the result indicates a genuine quantum correlation between coherent states with a mutual injection path without excess noise.

\subsection{Distribution functions for quadrature amplitudes}
\begin{figure}[htbp]
\begin{center}
\includegraphics[width=8cm,trim=0 0 0 0,clip]{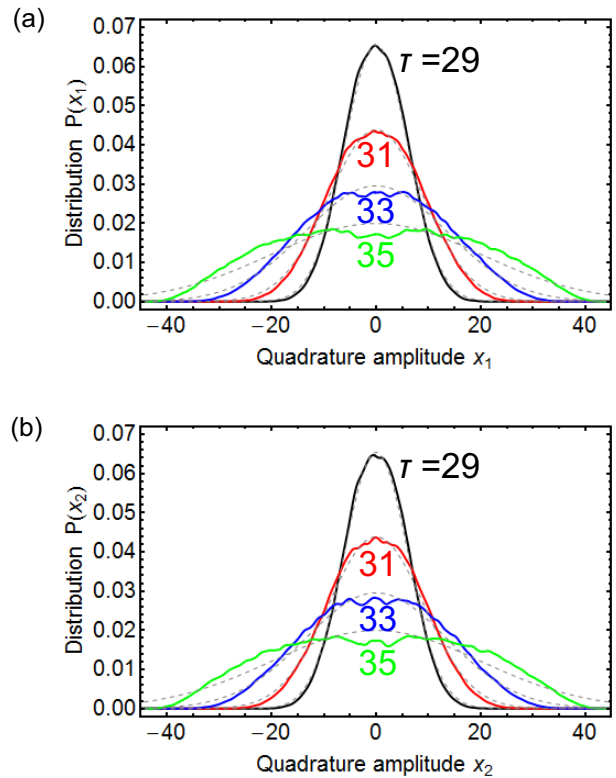} 
\end{center}
\caption{(Color online) Distribution functions at different time points for (a) $x_1$ and (b) $x_2$. The dashed lines are Gaussian fitting curves with $\sigma =$ (6.6, 9.1, 13.5, 20.0) for $\tau =$ (29, 31, 33, 35). 200,000 trajectories are used. $\gamma _s = 0.1$, $\gamma _c = 0.2$ and $g = 0.01$.}\label{fig:xDist2DOPO}
\end{figure}

\begin{figure}[htbp]
\begin{center}
\includegraphics[width=8cm,trim=0 0 0 0,clip]{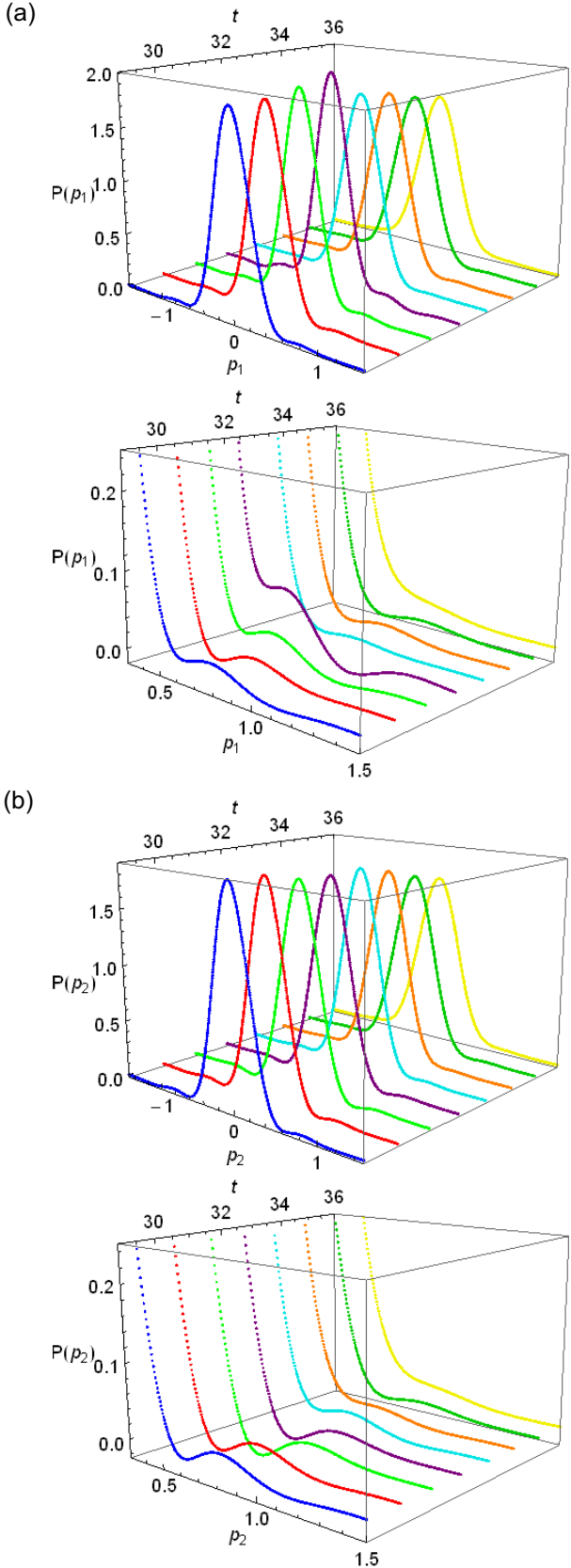} 
\end{center}
\caption{(Color online) Distribution functions at different time points for (a) $p_1$ and (b) $p_2$. Zoomed curves around the side peaks are added for both. 200,000 trajectories are used. $\gamma _s = 0.1$, $\gamma _c = 0.2$ and $g = 0.01$.}\label{fig:pDist2DOPO}
\end{figure}

Here, we focus on a low-loss case, where the distribution functions for the squeezed amplitude $p$ around the oscillation threshold are off from Gaussian curves. Fig. \ref{fig:xDist2DOPO} and \ref{fig:pDist2DOPO} displays instantaneous distributions for $x$ and $p$ at some time points for $\gamma _s = 0.1$, $\gamma _c = 0.2$. In Fig. \ref{fig:xDist2DOPO}, the distribution for $x$ gets broadened as the pumping rate increases. The dashed lines are Gaussian fitting curves for each time point. We see that it has some deviation from the fitting curve at $\tau = 33$ and $35$. This indicates that the system is at the onset of the macroscopic bifurcation in $x$.

As shown in Fig. \ref{fig:pDist2DOPO}, both $P(p_1)$ and $P(p_2)$ at pumping rates around the oscillation threshold come to have small fringes at the sides of their central peaks. The fringes survive until the clear bifurcation in $P(x_1)$ and $P(x_2)$. On the other hand, they vanish when $\gamma _s$ and $\gamma _c$ are comparable to or larger than $\zeta = 1$. Therefore, $P(p_1)$ and $P(p_2)$ suggests the existence of the macroscopic superposition of the zero-phase state and $\pi$-phase state in a sufficiently closed two-DOPO system. Formation of the superposition in such a slow pumping schedule with only a small $\gamma _c$ means that the quantum noise stored in the injection path is essential in the formation of superposition components here. The injection path contains a squeezed field for the two oscillators, which protects a macroscopic superposition state from decoherence \cite{paper:KW88,paper:MR95}. It is worth noting that the theoretical model considered here is different from that in the previous studies. The side peaks in $P(p_1)$ and $P(p_2)$ are as high as those in an even cat state $|-\alpha \rangle + |\alpha \rangle$ with $\alpha \sim 0.9$ although the state has a larger photon number than $|\alpha|^2 = 0.81$. This is because such a DOPO state does not correspond to a pure cat state. Note that the fringe signal will be a bit weaker than the flying optical cat states made with judicious techniques \cite{paper:GGCC10,paper:EBKT15}. Also, a larger $g$ and a faster pumping schedule will give a clearer fringe due to the transient effect, as the single DOPO case \cite{paper:KMR94}. 

\begin{figure}[htbp]
\begin{center}
\includegraphics[width=8cm,trim=0 0 0 0,clip]{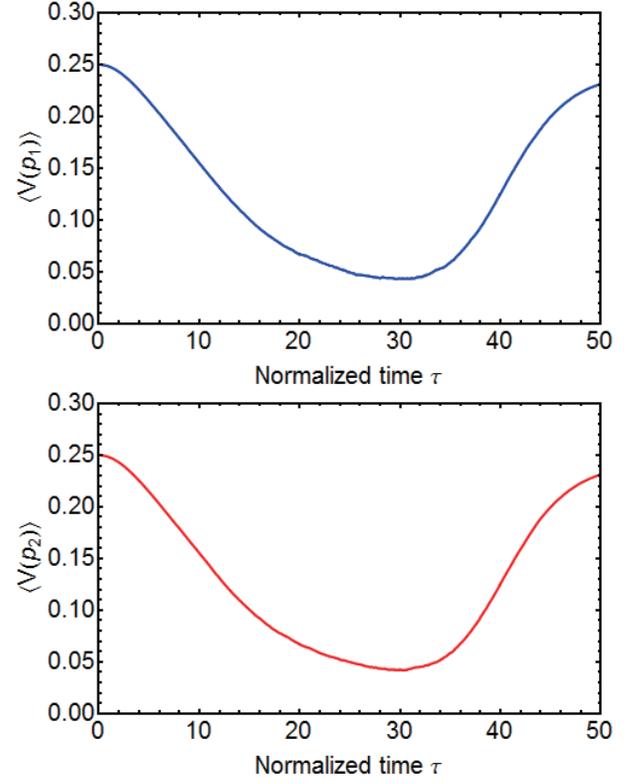} 
\end{center}
\caption{(Color online) Variances of $p_1$ (blue curve) and $p_2$ (red curve). The two curves are almost identical due to the same form of the SDEs for each DOPO. 200,000 trajectories are used. $\gamma _s = 0.1$, $\gamma _c = 0.2$ and $g = 0.01$.}\label{fig:VP2DOPO}
\end{figure}

Here, we add the extra squeezing of the intracavity DOPO fields which supports the effect by the mutual injections. Fig. \ref{fig:VP2DOPO} displays the variances of $p_1$ and $p_2$ versus time (i.e. the pumping rate). When the system is below the threshold, they decrease with the rise in the pumping rate. Following the oscillation of the DOPOs, they get back to the value for a coherent state and the vacuum state (0.25). The minimum value $\sim 0.043$ is smaller than that for a single intracavity DOPO field \cite{paper:CG84} (0.125, meaning -3 dB squeezing). It suggests that the mutual injections enhance the squeezing in the DOPOs.

Compared to expectation values of observables, the convergence of the distribution functions (Fig. \ref{fig:pDist2DOPO}) to the number of sampling is slower, because the sampled points have to cover the whole space where the distribution can have a non-negligible value. Thus, we have taken 200,000 runs to draw the curves here. Simultaneous formation of the side peaks in both $P(p_1)$ and $P(p_2)$ is a good indicator that the accuracy is not bad, because the two DOPOs obey the SDEs of the same form. However, numerical errors still lead to obvious negative values in some curves. Also, one of the $p$ distribution functions is fluctuated a lot at some time points, leading to a larger fringe visibility and negative values.

\section{Discussion}\label{sec:DC2DOPOMI}
In this section, we discuss other theoretical schemes to simulate the system considered here, the validity of the simulation in this study, and the possible contribution of the quantum effects in the system to the performance of the coherent Ising machines. 
\subsection{Other theoretical schemes}
First, we refer to the difficulty in the simulation in this study with other theoretical schemes. Regarding a numerical analysis for an open quantum system, direct integration on the master equation with the Fock state basis is the most standard method as investigated in the previous relevant studies \cite{paper:KW88,paper:MR95}. It treats a series of ordinary differential equations for the components of the density matrix for the system. Single-shot numerical integration for them gives all the information of the solution, thus we do not have to repeat stochastic simulations or take ensemble averages over a number of samples. Also, it is relatively easy to get a good accuracy in numerical integration of an ordinary differential equation. However, the basis has an infinite number of eigenstates hence we have to truncate some of them. Here, the more photons possible in the system, the more eigenstates needed. In addition, the number of modes crucially affects the complexity of the simulation. When we consider two DOPOs and the injection path with $m$ eigenstates for each, the number of components of the density matrix is $m^6$. This amounts to unrealistic numbers such as $1000^6$ and $10000^6$ thus the simulation with the parameters here will be too costly.

Solving the Fokker-Planck equation will be useful if we can find a potential solution. However, it supposes a system at the steady state thus cannot treat the transient regime, which is thought to be important for a DOPO to have macroscopic superposition components \cite{paper:RY92,paper:KMR94}. Also, when the mutual injection path is explicitly considered, we will not be able to find a potential solution.

The linearized analysis has been used to investigate the output squeezing spectra for the quadrature amplitudes of a DOPO \cite{book:WM08} and the entanglement between the output fields of two DOPOs coupled with evanescent coupling \cite{paper:OD05,paper:OO06}. We can apply this method to get some information about the intracavity fields of the DOPOs. However, this is also limited in the steady state, and basically gives a result in the frequency domain. To consider the properties in the time domain, we have to integrate the noise spectra over the entire frequency space. Here, a well-known relation between the covariance of input and output bosonic operators \cite{paper:CG84} is not satisfied. It is because the input field for each DOPO originating from the mutual injections is not a coherent or vacuum state but a squeezed state. Straightforward application of this relation results in negative values in the spectrum of $\langle \Delta v_{-}^2 \rangle$. Thus, reconsideration on the theoretical model might be required.

\subsection{Accuracy and limit of simulation}
Next, we discuss the stability of the simulation in this study. We recognize that the positive $P$ representation can give unreliable results in some cases \cite{paper:GGD97}. However, the simulation here is considered to be relatively stable, because the dynamics of the variables are mostly bounded in a finite manifold. In the theoretical model, the pump modes are adiabatically eliminated. This significantly helps DOPO fields be bounded as Ref. \citenum{paper:GGD97} pointed. It is also supported by the fact that the Fokker-Planck equation for a single DOPO with the adiabatic elimination has the solution \cite{paper:DMW81} which comprises Gaussian components decaying exponentially in the phase space. In addition, a real and diagonal diffusion amplitude matrix assists them to be real. The classical solution and the gradual pumping scheme also help the variables be bounded. Thus, the simulations here are not thought to be largely affected by the instability of the dynamics in the complex phase space.

Nevertheless, the mutual injection path can enhance the numerical error of the simulation, especially when its loss $\gamma _c$ is quite small. As mentioned, a relaxation oscillation might be a sign for the unreliability of the simulation. We have not seen large variance in the amplified quantities such as the photon number, the correlation, and fluctuation for $x$. However, those for the squeezed amplitude $p$ directly reflects the feature of the noise in the system. Thus, it can be difficult to acquire a good accuracy for them. We see that the curves for them with $\gamma _s = 0.05$ in the first case and $\gamma _c = 0.1$ in the second have large fluctuations, around which the simulation might be unreliable \cite{paper:GGD97}. Thus, we have taken $\gamma _s = 0.1$, $\gamma _c = 0.2$ and 200,000 samples for the distribution functions to avoid large numerical errors. The critical fluctuations around the oscillation can also be another difficulty because the range of the variables here is big due to a small $g$. However, it is worth noting that the noise in $p$ is rather decreased there, and that in $x$ does not diverge in principle because the nonlinear pump depletion is fully taken into account here. To reduce the number of samples by removing unexpected correlation in the noise terms and the instability, the introduction of the gauge terms \cite{paper:DD02} might be needed.

\subsection{Impact on coherent Ising machine}
Finally, we comment on the possibilities that the quantum features in the DOPOs contribute to the performance of the coherent Ising machine, which is a DOPO network with mutual injections emulating the Ising model. We have seen different quantum effects for different parameter regimes. First, the system shows transient macroscopic superposition around the oscillation threshold when it has a small loss compared to the oscillator mirror loss $\zeta$. The coherent Ising machine assigns the up and down spin states to the discrete phase states $|0\rangle$ and $|\pi \rangle$. It is expected to oscillate in a phase configuration with the small total gain, which corresponds to the ground state of the Ising Hamiltonian. For an efficient ground-state search, the network has to avoid being trapped in local minima, whose number scales exponentially with the problem size in hard instances. The macroscopic superposition under gradual pumping holds the information of all the states in the whole system simultaneously and can potentially help the system search for the minimum gain mode at the onset of the oscillation. This is similar to the quantum parallelism in quantum computers. Second, a loss of the mutual injections comparable to $\zeta$ leads to the quantum correlation in terms of the squeezed amplitude $p$. Further investigations are required to clarify its effect on the performance of the machine.
\section{Conclusion}\label{sec:CL2DOPOMI}
We have founded and simulated a fully quantum mechanical model of the system of two DOPOs with the mutual injections. The field between the two DOPO facets is introduced as a cavity mode to treat the input-output relation and the field confinement in the mutual injection path. The model can be extended easily to the case of a larger network. We have shown that the linear mutual injection terms in the positive $P$ representation are derived \textit{ab initio} in the limit where the dynamics of the injection path is neglected. The detailed simulation results for the case of out-of-phase coupling revealed that the gradually pumped system could have quantum correlation, entanglement and macroscopic superposition components. The quantum correlation and entanglement require noise-free mutual injections with moderate loss in its path. On the other hand, additional quantum noise stored in the low-loss injection path is rather essential to generate macroscopic superposition components in the DOPOs. This means that the closed injection path with squeezed vacuum inputs alleviates decoherence in the DOPOs. Quantum effects in such a simple setup may also play a role not only in a coherent Ising machine but also in broader contexts such as nano- and opto-mechanics, circuit QED and superconducting devices.
\begin{acknowledgments}
We thank H. Tajima, Y. Nakamura and Z. Wang for fruitful discussions. Also, we appreciate S. Pirandola and Y. Huang for their personal communications and comments about this paper. This work is supported by JST through its ImPACT program. K.T. thanks for Grant-in-Aid for JSPS Fellows.
\end{acknowledgments}

\end{document}